\documentclass[10pt]{iopart}
\usepackage{epsfig}
\newcommand{\be}{\begin{equation}}
\newcommand{\ee}{\end{equation}}
\newcommand{\tw}{t_{\rm w}}
\newcommand{\ta}{\tau_\alpha}
\newcommand{\tdh}{\tau_{\rm dh}}
\begin{document}

\title[Lifetime of dynamic heterogeneity 
in spin facilitated models]
{Lifetime of dynamic heterogeneity in strong and 
fragile kinetically constrained spin models}

\author{S\'ebastien L\'eonard and Ludovic 
Berthier\footnote[2]{To whom correspondence should be addressed:
berthier@lcvn.univ-montp2.fr}}

\address{Laboratoire des Collo\"{\i}des, Verres et Nanomat\'eriaux,
UMR 5587 CNRS and  Universit\'e Montpellier II,
34095 Montpellier Cedex 5, France}

\begin{abstract}
Kinetically constrained spin models are schematic
coarse-grained models for the glass transition 
which represent an efficient theoretical tool to study 
detailed spatio-temporal aspects of dynamic
heterogeneity in supercooled liquids.
Here, we study how spatially correlated 
dynamic domains evolve with time and compare our results
to various experimental and numerical investigations.
We find that strong and fragile models yield different 
results. In particular, the lifetime of dynamic heterogeneity 
remains constant and roughly equal 
to the alpha relaxation time in strong models, while 
it increases more rapidly in fragile models
when the glass transition is approached.
\end{abstract}

\section{Introduction}
\label{Intro}

The viscosity of supercooled liquids increases
extremely rapidly when the temperature is reduced towards
the glass temperature. It is firmly established that 
this dramatic slowing down is spatially heterogeneous.
Local relaxation is widely distributed
in time --- existence of broad stretched relaxations ---, but
also in space --- existence of dynamic heterogeneity~\cite{ediger}.
The main physical aspect is that spatial 
fluctuations of local relaxations become increasingly spatially
correlated when temperature decreases. Direct
experimental investigations of the time and temperatures 
dependences of the relevant dynamic lengthscales at low temperature 
are however still missing.

To study dynamic heterogeneity, statistical correlators which probe
more than two points in space and time have to be 
considered~\cite{ediger,sillescu,glotzer}.
For example, if one wants to study spatial correlations
of the local dynamics one has to define a two-point, 
two-time correlator, 
\be
\label{22}
C_{2,2}(|i-j|,t) = \langle P_i(0, t) P_j(0, t) \rangle
-\langle P_i(0, t) \rangle \langle P_j(0, t) \rangle,
\ee  
where notations are adapted to 
lattice spin models. In equation~(\ref{22}), $(i, j)$ denote 
lattice sites, $P_i(0, t)$ quantifies
the dynamics at site $i$ between times 0 and $t$ (autocorrelation
or persistence functions), and brackets represent ensemble averages.
The physical meaning of (\ref{22}) is clear: given a spontaneous 
fluctuation of the two-time dynamics at site $i$, is there a similar 
fluctuation at site $j$? The quantity (\ref{22}) has now been 
discussed both theoretically and numerically in some 
detail~\cite{glotzer,glotzer2,wyart}, 
generically revealing the existence of a growing spatial range 
of dynamic correlations in supercooled liquids accompanying 
an increasingly sluggish dynamics.

Logically, the next question is then: given spatial structures
of the local 
relaxation between times 0 and $t$, what will this structure look like,
say, between times $t$ and $2t$? In other words~\cite{heuer1}, 
how do dynamic heterogeneities evolve with time?
This question is in fact simpler to address experimentally
because no spatial resolution is needed 
and different experimental techniques can be devised:
NMR~\cite{nmr}, solvation dynamics~\cite{solvation}, 
optical~\cite{ediger3} and dielectric~\cite{hole} hole-burning.
In statistical terms, one wants to study
a four-time correlation function of the general form
\be
C_4(t_1, \tw, t_2) = \langle P_i(0, t_1) P_i(t_1+\tw, t_1+\tw+t_2) \rangle, 
\label{4}
\ee
which correlates dynamics between times 0 and $t_1$
and between times $t_1+\tw$ and $t_1+\tw+t_2$. Again the 
physical content of (\ref{4}) is clear~\cite{heuer2}: given a dynamic
fluctuation at site $i$ between in a certain time interval $t_1$, 
how long does it take for this fluctuation to be washed out? 
This leads to the general concept of a lifetime, $\tdh$, for dynamic
heterogeneity. While many 
investigations~\cite{ediger,sillescu,nmr,solvation,hole,harrowell,szamel} 
indicate that $\tdh$
is in fact slaved to the alpha-relaxation time of the 
liquid, $\tdh \approx \ta$, photobleaching experiments 
very close to the glass transition
indicate  that $\tdh$ may become several 
orders of magnitude larger than $\ta$, although in a
surprisingly abrupt manner~\cite{ediger2}. 

In this paper we study the lifetime of dynamic heterogeneity
in kinetically constrained spin models of supercooled 
liquids~\cite{reviewkcm}. 
These models represent schematic coarse-grained models 
for the glass transition and provide a very efficient tool to study 
in detail many spatio-temporal aspects related to dynamic
heterogeneity such as  dynamic lengthscales~\cite{gc}, 
scaling~\cite{steve}, or decoupling 
phenomena~\cite{jung,epl}. 
They are simple enough that analytical 
progress can be made and numerical simulations performed 
on a wide range of lengthscales and timescales, and yet 
rich enough that direct comparisons to both simulations
and experiments can be made.

\section{Models} 

Following previous works~\cite{gc,jung,epl,BG}, 
we focus on two specific 
spin facilitated models in one spatial dimension, namely 
the one-spin facilitated 
Fredrickson-Andersen (FA) model~\cite{fa}
and the East model~\cite{east} that respectively behave as strong 
and fragile systems~\cite{reviewkcm}. 
These are probably 
the simplest models which  incorporate the ideas that (i) mobility
in supercooled liquids is both highly localized and sparse, as
revealed by simulations~\cite{harrowell}; 
(ii) a localized mobility very 
easily propagates to neighbouring regions, the 
dynamic facilitation concept. 
Detailed studies in spatial dimensions larger than one have shown
that dimensionality does not play a relevant qualitative 
role~\cite{steve,jung,nef},
and justify therefore the present one-dimensional studies. 

Both models are defined by the same non-interacting Hamiltonian, 
$H = \sum_i n_i$, expressed in terms of a mobility variable, $n_i=1$
when site $i$ is mobile, $n_i=0$ otherwise.
Dynamic facilitation is incorporated at the level of the 
dynamic rules through kinetic constraints. 
In the FA model, the site $i$ can evolve with Boltzmann 
probability if at least one of its two neighbours is mobile, 
$n_{i-1}+n_{i+1} > 0$. 
In the East model the site $i$ can evolve only if its left 
neigbour is mobile, $n_{i-1}=1$.

We have performed numerical simulations of both models
using a continuous time Monte Carlo algorithm where all moves
are accepted and the time is updated according to the 
corresponding statistical 
weight. Simulations have been performed 
`only' over about 7 decades in time because 
extensive time averaging is required to accurately 
measure multi-time correlation functions such as equation~(\ref{4}).

\section{Results}

\subsection{Dynamic filtering}

There are several parameters involved in the four-time 
correlator (\ref{4}) that need to be appropriately chosen.
Since dynamic heterogeneity is more pronounced for times 
close to $\ta$ it is sensible to first fix 
$t_1 = \ta$ and to study the remaining $\tw$ 
and $t_2$ dependences.
As a local dynamic correlator we first
focus on the persistence function~\cite{heuer5}, 
$P_i(0,t)=1$ if spin $i$ has not flipped in the interval $[0,t]$, 
$P_i(0,t)=0$ otherwise. 
We also define the mean persistence, 
$p(t) = \langle P_i(t) \rangle$, from which we measure 
$\ta$ via $p(\ta)=e^{-1}$.   

\begin{figure}
\begin{center}
\psfig{file=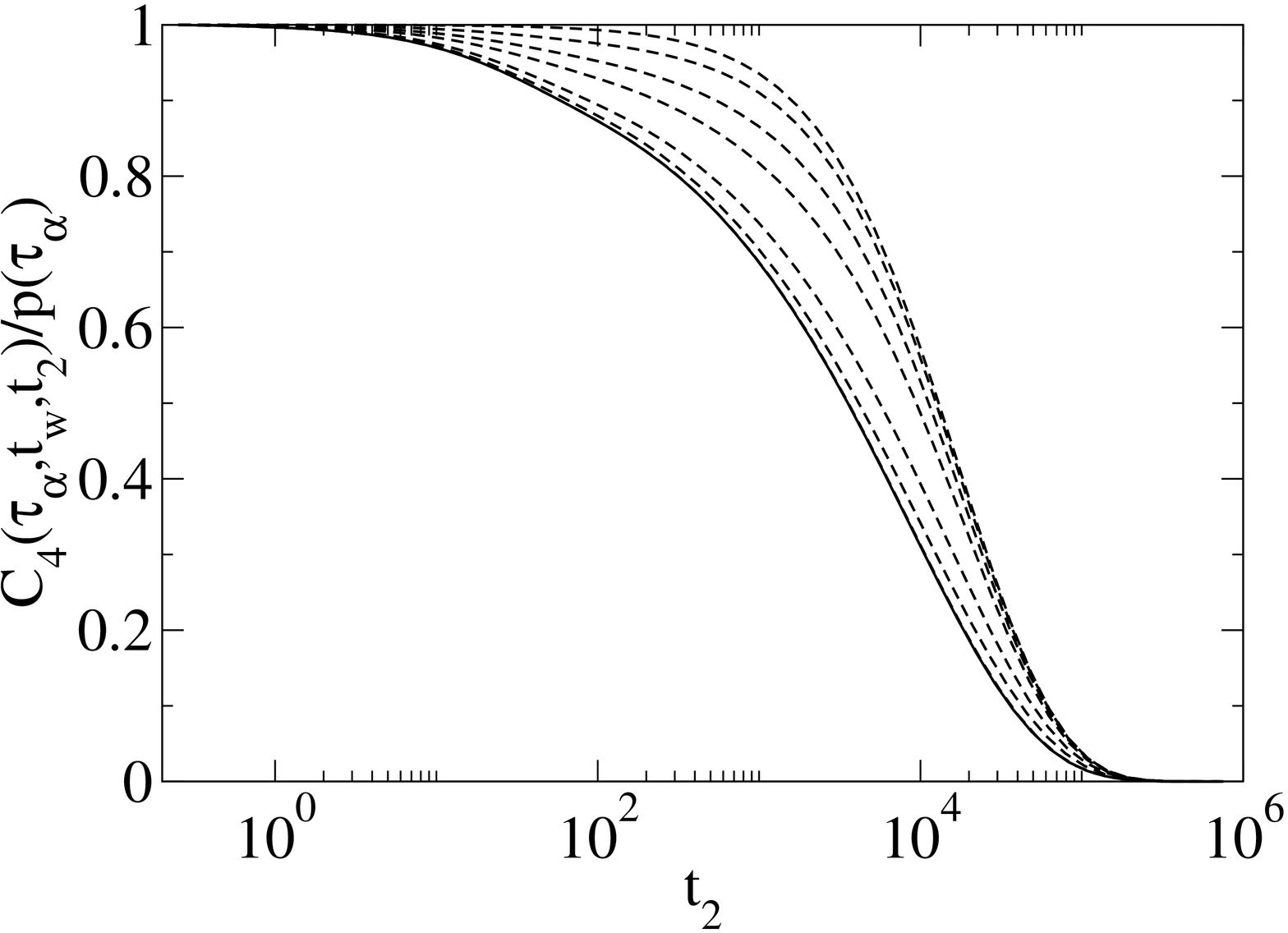,height=6.45cm,angle=-90}
\psfig{file=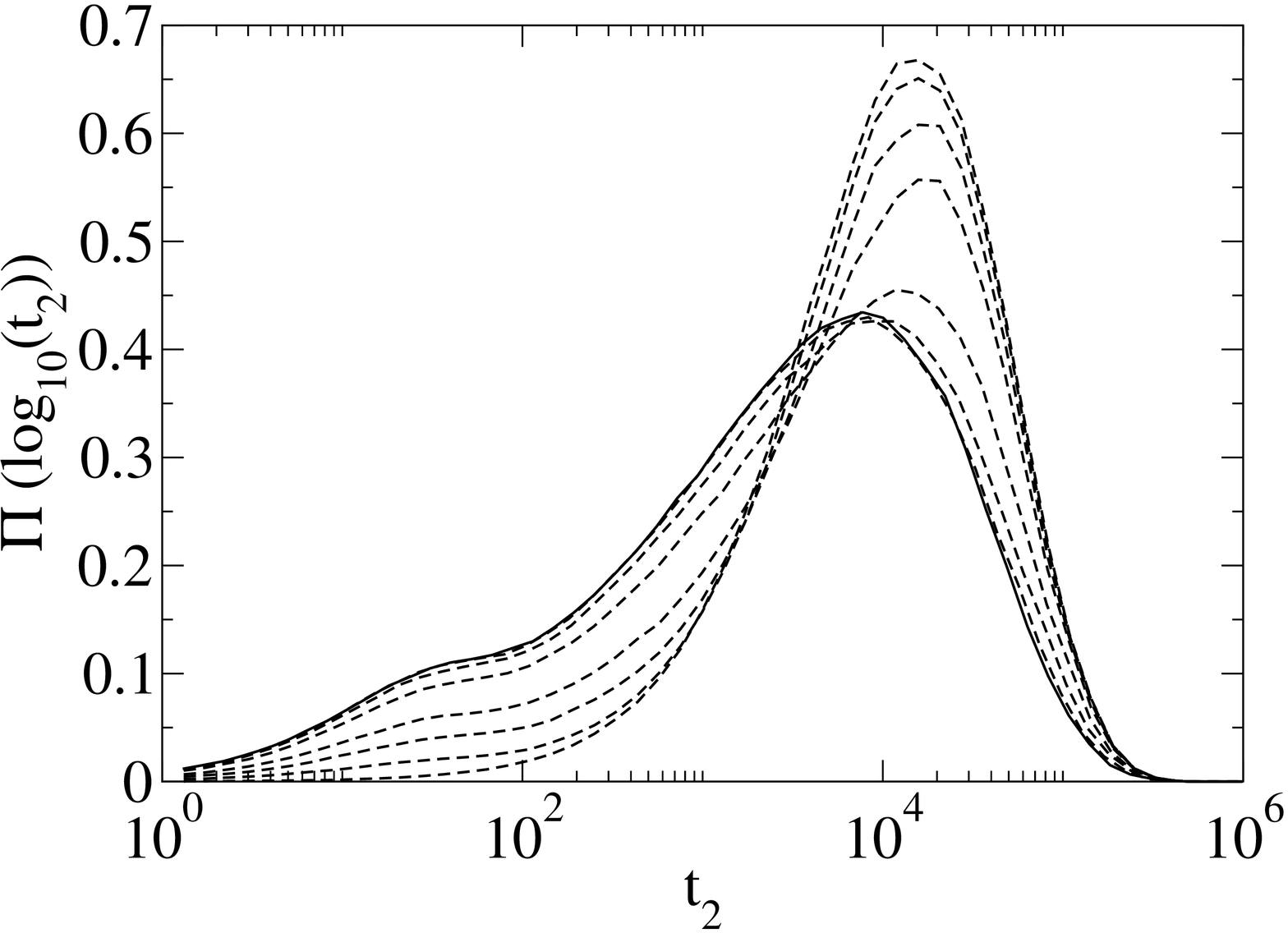,height=6.45cm,angle=-90}
\caption{\label{fig1}  Left: Four-time correlation
function $C_4(\tau_\alpha, \tw, t_2)$ normalized by its 
$t_2=0$ value for various $\tw=0$, 
550, 2100, 5200, 20000, 49500, and 191000 (from right to left)
in the FA model at $T=0.3$. These dashed curves converge 
at large $\tw$ to the bulk persistence function, $p(t_2)$,  
shown as a full line.
Right: Corresponding logarithmic distributions of relaxation 
times.}
\end{center}
\end{figure}

In \fref{fig1} (left) we show the $t_2$ dependence
of $C_4(\tau_\alpha, \tw, t_2)$ for various $\tw$
at $T=0.3$ in the FA model. We have normalized 
$C_4$ by $p(\ta)=e^{-1}$, its value at $t_2=0$.
By definition, this function describes the persistence 
function in the interval $[\tw+\ta, \tw+\ta+t_2]$ of those sites
which had not flipped in the interval $[0, \ta]$, 
and were therefore slower than 
average.
The first term in the correlator (\ref{4}) plays the role
of a dynamic filter~\cite{heuer2}, selecting a sub-population of sites
which have an average dynamics different from the bulk.
From earlier works studying the spatial correlator
(\ref{22}), it is known that those sites belong to 
compact clusters that represent the largest regions of space with 
no mobility defects at time 0~\cite{gc,steve,nef}.
 
Immediately after filtering one expects therefore those slow regions
to remain slow, as 
indeed observed in \fref{fig1} for $\tw=0$. 
When $\tw$ increases, this selected population gradually forgets
it was initially slow. When $\tw \to \infty$, bulk
dynamics is recovered,
\be 
\frac{C_4(t_1, \tw \to \infty, t_2)}{p(\ta)} \to p(t_2), 
\ee
as demonstrated by the full line in \fref{fig1}. 
In \fref{fig1} (right) we also show the 
(logarithmic) distribution 
of relaxation times corresponding to the functions shown 
in the left panel, a representation sometimes preferred 
in experimental works~\cite{ediger}. Both quantities are of course
fully equivalent~\cite{BG}. 
It is clear from these figures that once a subset 
of sites has been dynamically selected the remaining relaxation
is narrower than the bulk relaxation. In fact all persistence functions
shown in \fref{fig1} are well described by stretched 
exponentials. While $\beta=1/2$ is observed for the bulk dynamics, 
one finds $\beta \approx 0.83$ at $\tw=0$. Accordingly, distribution
of relaxation times progressively broaden 
when $\tw$ increases. These results are consistent with experimental 
observations.

In the FA model their 
interpretation is straightforward. Stretching in this model
follows from an exponential distribution of distance between mobility 
defects~\cite{gc}.
Dynamic filtering implies that this domain distribution is cut-off 
at small distance. Narrower lengthscale distributions 
directly imply narrower timescale distributions. 

We have also investigated the effect of changing the 
`filter efficiency'~\cite{nmr} which
in our case implies changing the duration of the filtering interval, 
$[0,t_1]$. While bulk distributions are found for 
$t_1/\ta \ll 1$ (weak filtering) 
distributions shift to larger times and become very narrow 
when $t_1/\ta$ increases. 
In the following we work at constant 
filtering, $t_1 = \ta$. 

\subsection{`Homogeneous' vs. `heterogeneous' dynamics}

The ability to select a sub-ensemble of sites that 
are slower than average is sometimes taken as a definition of 
dynamic heterogeneity~\cite{nmr}, although lengthscales play no role
in this view. That FA and East model display 
spatially heterogeneous dynamics is well-known, 
and the results of the previous section are therefore 
natural. 

\begin{figure}
\begin{center}
\psfig{file=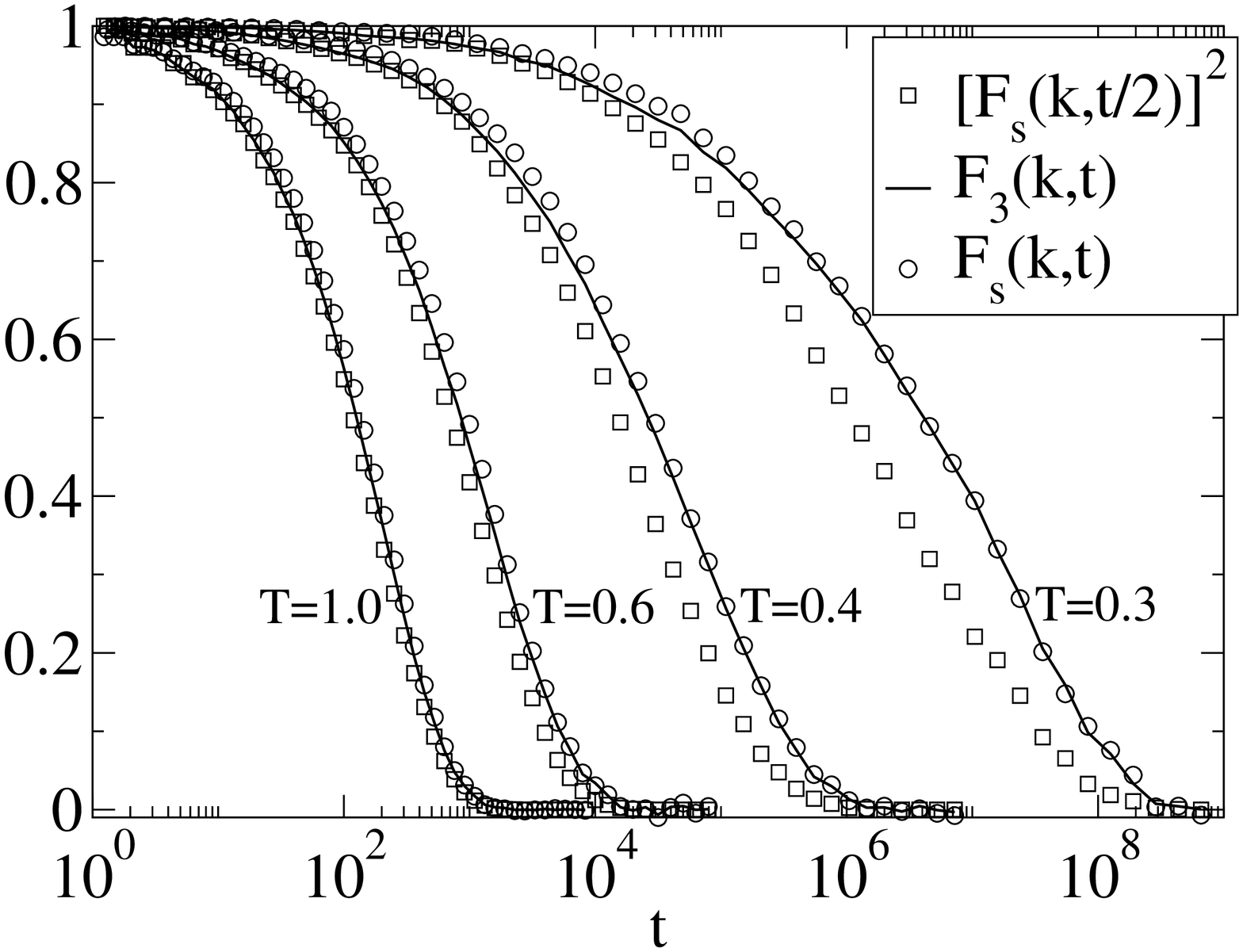,height=6.4cm,angle=-90}
\psfig{file=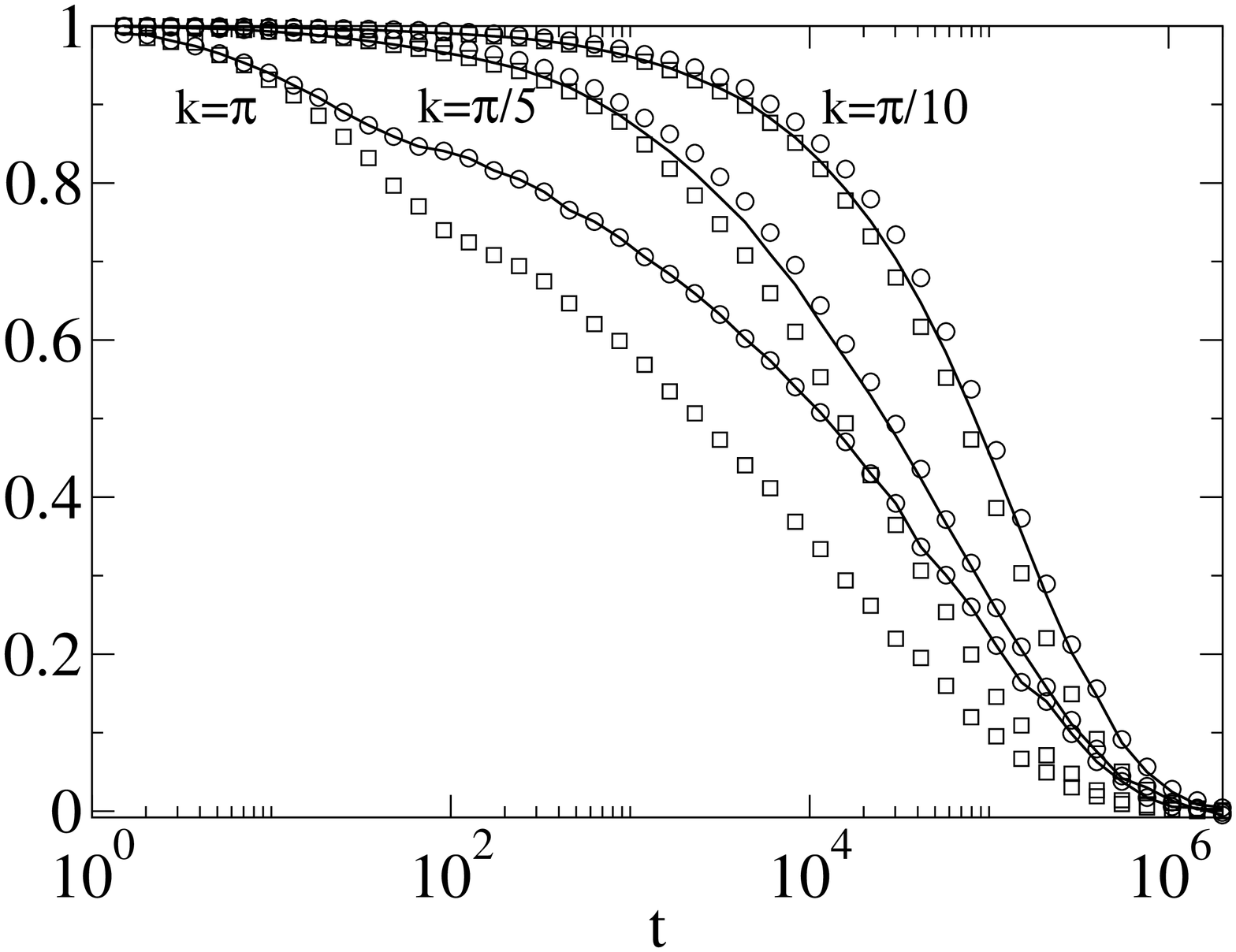,height=6.4cm,angle=-90}
\caption{\label{fig3} $F_3(k,t)$ and its `heterogeneous', $F_s(k,t)$,
and `homogeneous', $[F_s(k,t/2)]^2$, limits.
Left: East model for fixed $k=\pi/5$ and various temperatures. 
Right: East model at $T=0.4$ and various wavevectors.}  
\end{center}
\end{figure}

Another indicator of dynamic heterogeneity has been 
proposed~\cite{heuer2,heuer3,heuer4} 
based on the analysis of the four-time correlation
(\ref{4}). Consider the situation where $\tw=0$, and $t_1=t_2 \equiv t/2$. 
In that case, one studies a `three-time' correlation~\cite{heuer4}
\be 
F_3(t) = \langle P_i(0,t/2) P_i(t/2,t) \rangle.
\label{F4}
\ee 
Two extreme behaviours can be expected for $F_3(t)$. (i) Dynamics 
in the intervals $[0,t/2]$ and $[t/2,t]$ are totally uncorrelated, 
and thus $F_3(t) \approx [p(t/2)]^2$.
(ii) Dynamics in the two intervals are strongly correlated, in the sense 
that those regions that survive 
filtering in $[0,t/2]$ are also those dominating the relaxation
in the full interval $[0,t]$. In that case, $F_3(t) \approx p(t)$. 
Scenarii (i) and (ii) have been termed `homogeneous' and 
`heterogeneous', respectively, although again lengthscales play no
role in the distinction.
Clearly, both estimates become equivalent
when $p(t)$ decays exponentially.

Of course when studying the persistence function in the FA and East models, 
scenario (ii) strictly applies by definition, because 
$P_i(0,t/2) P_i(t/2,t) = P_i(0,t)$. In real materials,
smoother dynamic functions are studied, directly defined 
from the particles positions instead of a mobility field. 
Our strategy is therefore to couple probe particles to our 
mobility field, see Refs.~\cite{jung,epl} for technical details. 
From probe molecule displacements, $\delta x(0,t) = x(t)-x(0)$, 
we define self-intermediate scattering functions, $F_s(k,t) = 
\langle \cos[k \cdot \delta x (0,t) ] \rangle$, and the analog of 
equation (\ref{F4}), $F_3(k,t) = \langle \cos [k \cdot \delta x (0,t/2)] 
\cos [k \cdot \delta x (t/2,t)] \rangle$. 

Our numerical results are presented 
in \fref{fig3}.
Clearly the time dependence of $F_3$ closely 
follows the one of $F_s(k,t)$, in agreement with the `heterogeneous' scenario
described above. This is consistent with numerical
results~\cite{heuer4}. 

In the present approach, this result is a natural consequence
of decoupling between structural relaxation and 
diffusion~\cite{jung,epl}.
At large wavevectors, $k \sim \pi$, 
corresponding to distances of the order of the lattice 
spacing, $F_s(k,t)$ is dominated by the time distribution 
of the first jump of the probe molecule in the interval $[0,t]$, 
so that $F_s(k,t) \approx p(t) \approx F_3(k,t)$.
At large distance, $k < k^*$, Fickian diffusion holds~\cite{epl},
$F_s(k,t) = \exp(-k^2 D_s t)$, and there is no distinction between
homogeneous and heterogeneous relaxation.
At intermediate wavectors, $\pi > k > k^*$, 
the long-time decay of $F_s(k,t)$ is again
dominated by the persistence time distribution,
because the timescale it takes a molecule to make
$2 \pi / k$ steps is strongly dominated by the 
timescale to make the first step~\cite{epl}.
This is just the condition for the heterogeneous 
scenario to hold, in agreement with \fref{fig3}.
The characteristic wavevector separating the two regimes,
$k^*(T) = 1/\sqrt{\ta D_s}$, decreases when temperature decreases, 
opening a larger heterogeneous window; $k^*$ also 
sets the upper limit of validity of Fickian diffusion
in supercooled liquids~\cite{epl}.

\subsection{Lifetime of dynamic heterogeneity}

After dynamic filtering it takes some time for filtered
distributions to reequilibrate towards the bulk relaxation, 
cf \fref{fig1}.
To extract the typical lifetime of dynamic heterogeneity, $\tdh$, 
we tried several procedures which all lead to similar results, based
on how timescales (time decay of persistence functions
or moments of the corresponding distributions)
return to their equilibrium values.
Following Refs.~\cite{szamel,jung} we also measured the
integrated difference between filtered and bulk dynamics, 
$\Delta (\tw) \equiv \int_0^\infty dt_2 [
C_4(\ta, \tw, t_2)/p(\ta) - p(t_2)]$.
From \fref{fig1}, we expect that $\Delta (\tw)$ goes to 0
on a timescale $\tdh$. In practice, we define
$\tdh$ as $\Delta(\tdh)/\Delta(0) = e^{-1}$. 
In principle, $\tdh$ depends on the filtering time, $t_1$, and on
temperature, $T$. 

\begin{figure}
\begin{center}
\psfig{file=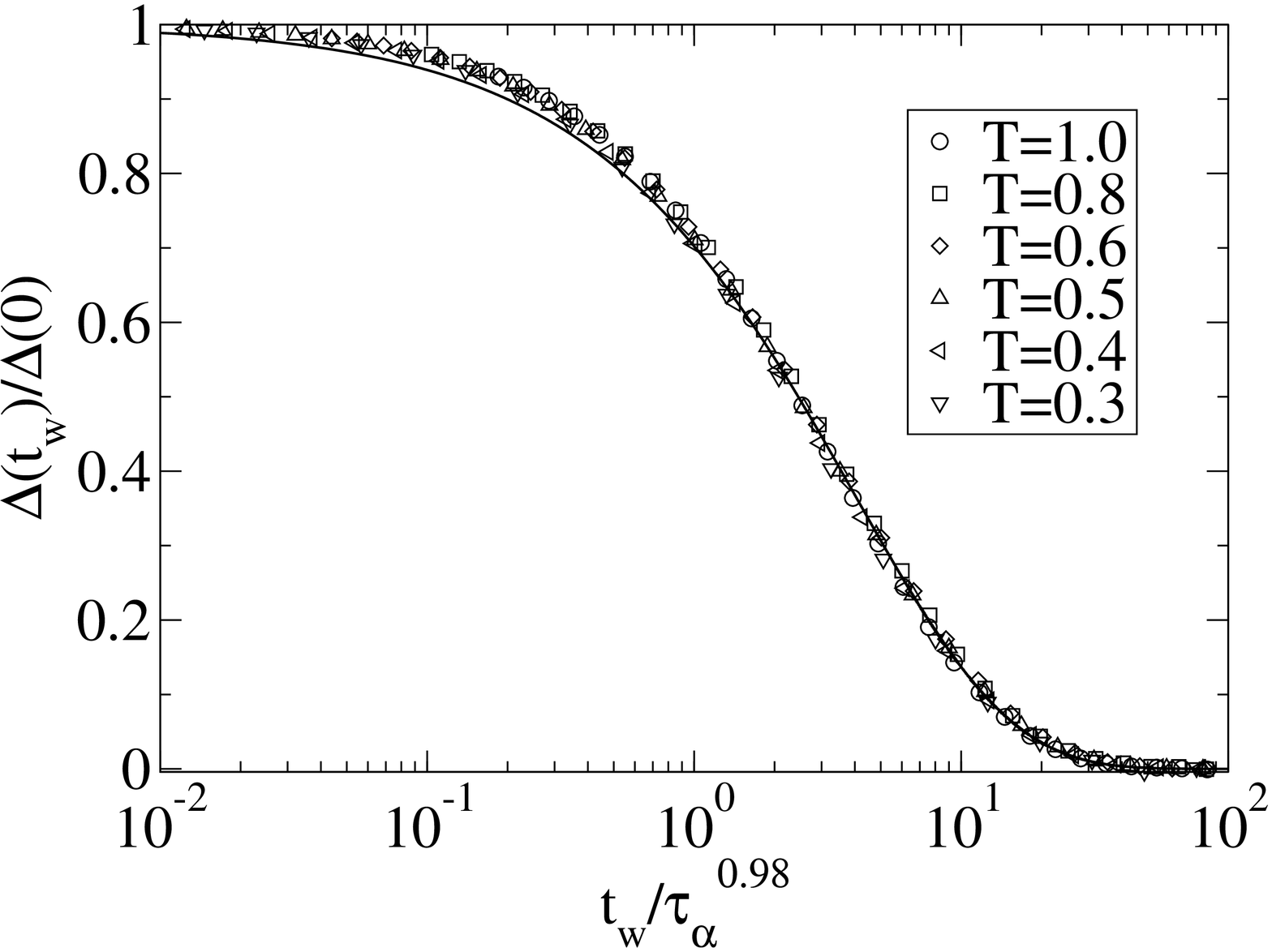,height=6.4cm,angle=-90}
\psfig{file=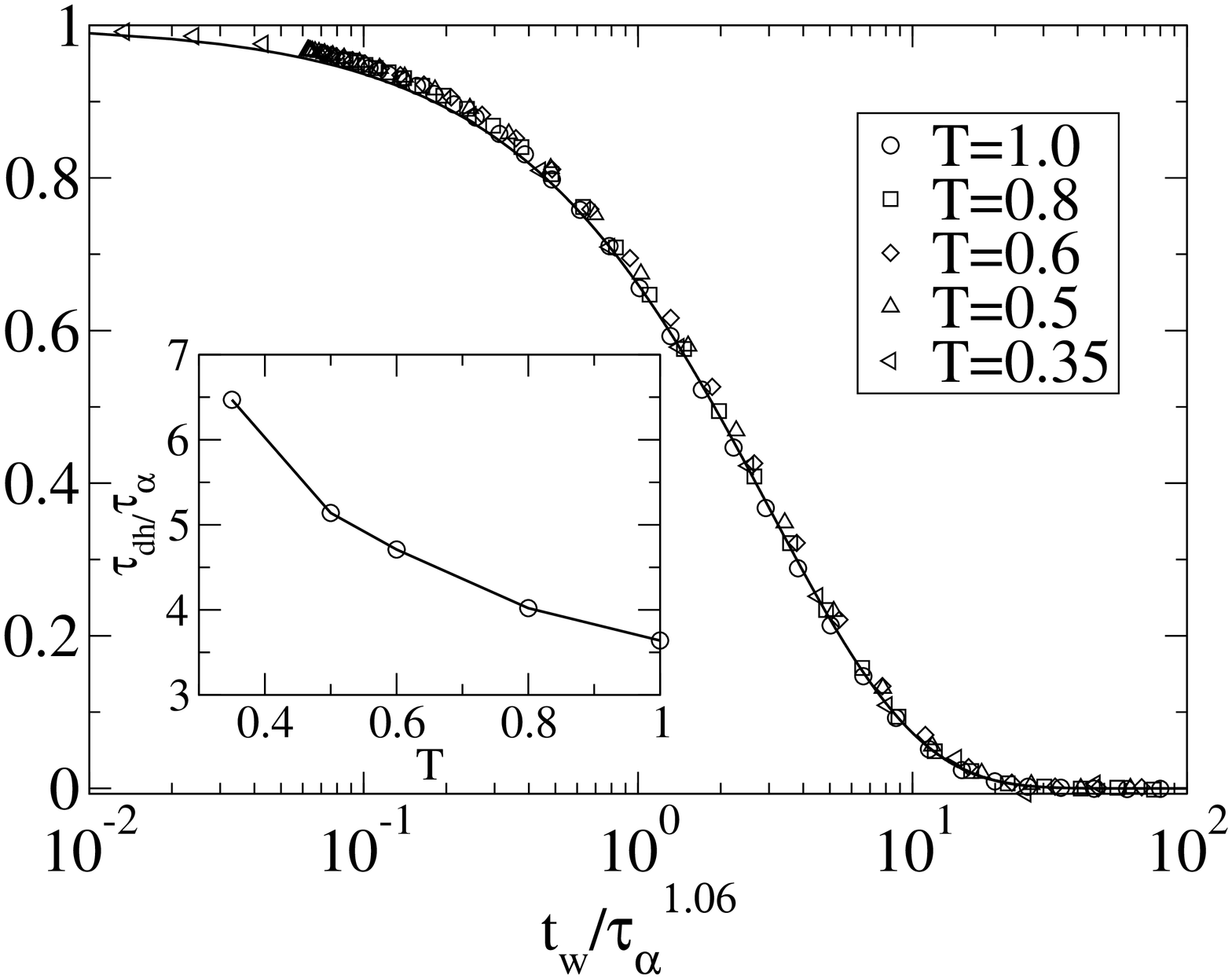,height=6.05cm,angle=-90}
\caption{\label{fig4} Integrated difference between 
filtered and bulk dynamics at constant filtering time, $t_1 = \ta$, 
and various temperatures in the FA (left) and East (right) models. 
Times have been rescaled to collapse the data points at various 
temperatures and extract the lifetime of dynamic heterogeneity, $\tdh$. 
In the FA model, $\tdh$ is set by $\ta$ with tiny corrections
that vanish at low $T$. In the East model, 
$\tdh$ grows faster than $\ta$, as emphasized in the inset
showing the systematic increase of $\tdh/\ta$ as $T$ decreases.
Mastercurves are fitted by stretched exponentials shown 
with full lines, $\beta=0.75$ (left) and $\beta=0.8$ (right).}  
\end{center}
\end{figure}

We show in \fref{fig4} results at 
various $T$ but constant 
filter efficiency, $t_1 = \tau_\alpha(T)$, in East 
and FA models. 
While $\tdh$ is set by $\ta$ in the FA model 
(the tiny deviation observed in \fref{fig4} 
is due to finite $T$ corrections which weaken when $T$ 
gets lower),
this is not true in the East model where $\tdh$ 
systematically grows faster than 
$\ta$ at low $T$, as emphasized in the inset.
Quantitatively a power law relationship, 
$\tdh \sim \ta^{1+\zeta}$, with $\zeta \approx 0.06$,  
is a good description of the data, although alternative 
fitting formula could probably be used. 

In the fragile case, $\tdh$ can therefore be considered as an additional 
slow timescale characterizing the 
alpha-relaxation~\cite{ediger2}, on top 
of $\ta$ and $1/D_s$~\cite{jung}. 
The comparative study of FA and East models
offers a possible physical interpretation. While both models
display stretched relaxations, in the FA model stretching is constant,
$\beta=1/2$, while $\beta$ increases linearly with $T$ in the East 
model~\cite{BG}. Therefore $\ta$ represents the first moment
of a distribution that becomes wider and wider 
when $T$ decreases. We attribute the small but systematic 
decoupling between $\tdh$ and $\ta$ to this broadening. 

Unfortunately this decoupling does not quantitatively account for 
the results of photobleaching experiments which show that 
$\tdh/\ta$ increases strongly close to $T_g$~\cite{ediger2}. 
In OTP, while 
$\ta$ changes by about 1 decade when $T$ is changed from $T_g+4$~K to
$T_g+1$~K, the ratio $\tdh/\ta$ changes by 2 orders of magnitude, 
so that $\zeta \approx 2$. This value is much too large 
to be accounted for by the above results. Presumably, also, $\beta$ does
not vary much on such a tiny temperature interval. 
Therefore, the present results cannot explain 
the experimental value $\zeta \approx 2$ without invoking 
possible non-equilibrium effects due to the proximity of 
$T_g$. However we were able to predict instead 
a smaller, but definitely non-vanishing decoupling between 
the lifetime of dynamic heterogeneity and 
the alpha-relaxation time which could be detected in 
dynamic filtering experiments performed on a sufficiently large temperature
window in fragile glass-formers.

\end{document}